# History of generative Artificial Intelligence (AI) chatbots: past, present, and future development

Md. Al-Amin[1], Mohammad Shazed Ali[2], Abdus Salam[3], Arif Khan[4], Ashraf Ali[5], Ahsan Ullah[6], Nur Alam[6], Shamsul Kabir Chowdhury[1]

[1]University of Massachusetts Lowell, MA, USA

[2]Asian University of Bangladesh, Dhaka, Bangladesh

[3]Sonargaon University, Dhaka, Bangladesh

[4]Babes-Bolyai University, Cluj-Napoca, Romania

[5]Islamic University, Kushtia, Bnagladesh

[6]University of Chittagong, Chittagong, Bangladesh

## Abstract

This research provides an in-depth comprehensive review of the progress of chatbot technology over time, from the initial basic systems relying on rules to today's advanced conversational bots powered by artificial intelligence. Spanning many decades, the paper explores the major milestones, innovations, and paradigm shifts that have driven the evolution of chatbots. Looking back at the very basic statistical model in 1906 via the early chatbots, such as ELIZA and ALICE in the 1960s and 1970s, the study traces key innovations leading to today's advanced conversational agents, such as ChatGPT and Google Bard. The study synthesizes insights from academic literature and industry sources to highlight crucial milestones, including the introduction of Turing tests, influential projects such as CALO, and recent transformer-based models. Tracing the path forward, the paper highlights how natural language processing and machine learning have been integrated into modern chatbots for more sophisticated capabilities. This chronological survey of the chatbot landscape provides a holistic reference to understand the technological and historical factors propelling conversational AI. By synthesizing learnings from this historical analysis, the research offers important context about the developmental trajectory of chatbots and their immense future potential across various field of application which could be the potential take ways for the respective research community and stakeholders.

1. **Introduction**

The term "chatbot," is a fusion of "chat" and "robot," originates from its initial function as a text-based dialogue system simulating human language. These early versions, primarily computer programs, employed input and output masks to create a mobile user experience mimicking a real-time conversation. However, chatbots have significantly evolved beyond this basic text-based interaction. Modern research delves deeper into their inner workings, exploring aspects, such as artificial intelligence (AI), natural language processing (NLP), and machine learning (ML). This multifaceted approach allows chatbots to not only respond but also learn and adapt over time, fostering more personalized and engaging interactions[1].

In the early era of chatbot development, pattern matching techniques were used to develop simple chatbots. It involved defining a set of rules and templates that map input phrases/sentences to appropriate responses. The chatbot searches the input for predefined keywords, structures, or patterns. If a match is found, the chatbot responds with the corresponding output template. These templates can contain static text, variables, or processing logic. Eliza and ALICE, the pioneering chatbots, were early examples employing pattern recognition algorithms [2]. The system identifies the longest matched pattern, deeming it the most precise, and selects the corresponding template as the output. Template-based chatbots emerge as a favored choice for numerous small-scale applications due to their simplicity in creation and configuration. It is useful while making small talk that needs low computing configuration and the ability to code domain-specific conversational rules. A wide-ranging topic is difficult for these chatbots. Their capabilities are restricted by the templates created by developers [3]. However, this approach has limitations, as the responses tend to be predictable, repetitive, and lack a human touch, and there is no retention of past responses, potentially resulting in repetitive and looping conversations[2]. Thus, the pattern matching with templates formed the early foundations of chatbot development.

The development of the Artificial Intelligence Markup Language (AIML) occurred between 1995 and 2000, centered around the principles of Pattern Recognition or Pattern Matching. This XML-based markup language is tag-oriented, specifically designed for natural language modeling in human-chatbot interactions, employing a stimulus-response framework [2]. AIML, an XML-based language tailored for chatbot development, has gained prominence as a widely used tool for creating conversational agents. With AIML, developers can craft rules that dictate the chatbot's responses to distinct user inputs, enabling natural and engaging human-computer interactions. The language's versatility and adaptability have positioned it as a fitting option for constructing chatbots across diverse contexts, including applications in mental health care[4].

In the early 2000s, the field of machine learning experienced significant advancement with the introduction of deep learning, enabling computers to comprehend and interpret information in various forms, such as text, images, audio, and videos. Major technology corporations actively drove the progress of artificial intelligence, harnessing the computational power to address complex challenges. The increased confidence in machine learning algorithms, bolstered by substantial datasets, facilitated their deployment in production environments, enhancing real user experiences. This transition from theoretical problems to practical implementation has empowered internet companies to utilize machine learning effectively, with leading tech firms contributing by making these algorithms openly accessible for widespread application and innovation [5].

In the early 2020s, the AI chatbot market is witnessing rapid expansion with the introduction of Bard and ChatGPT. Both these platforms, now available to the public, leverage the Transformer Neural Network Architecture, a research area that merits further exploration due to its distinctive characteristics and advantages. Originating from Google researchers, the Transformer Architecture presents a novel approach in neural networks, particularly beneficial for natural language processing[6]. Its superiority becomes evident in handling long and complex input sequences, offering enhanced performance through selective attention to different segments, making it invaluable for tasks like language generation and comprehension. The training process involves exposing the model to an extensive corpus of textual data, enabling it to recognize language patterns and correlations. This makes the Transformer Architecture well-suited for training language models such as ChatGPT and Bard, emphasizing their proficiency in generating coherent and contextually appropriate responses [7].

Numerous choices, both commercial and open-source, are accessible for chatbot development, contributing to the expanding array of chatbot-related technologies. The development of chatbots occurs through two primary approaches: utilizing programming languages such as Java, Clojure, Python, C++, PHP, Ruby, and Lisp, or employing state-of-the-art platforms. The abundance of available options underscores the dynamic and continually evolving landscape of chatbot technology[2]. Leveraging hyper-focused Customer Experience (CX) data alongside Artificial Intelligence can revolutionize the role of chatbots in customer support, moving beyond conventional rules-based methods to deliver more personalized and empathetic interactions. Although traditional chatbots remain suitable for specific workflows, the potential of AI-driven bots to comprehend user intent and facilitate dynamic, engaging conversations presents a significant opportunity for elevating the overall customer experience. As this technology progresses, organizations are encouraged to anticipate the future landscape and assess how generative AI can seamlessly integrate into and optimize their customer support strategies[8].

As chatbots continue to advance, their integration into various spheres of our lives, ranging from healthcare to education, is anticipated. Nevertheless, there are apprehensions regarding the potential effects of AI-powered chatbots on employment and privacy. Consequently, developers and policymakers must conscientiously examine the ethical ramifications of this technology as it progresses, emphasizing the need for thoughtful consideration and responsible implementation[9]. While there are scattered reports on the evolution of chatbots, a comprehensive review tracing the major milestones from the pioneering days of ELIZA and PARRY to the era of intelligent conversational agents is lacking. In particular, the interplay between advances in linguistic processing, machine learning, and the evolving capabilities of chatbots over time has not been thoroughly documented. This highlights a gap in research literature surveying the complete arc of chatbot technologies and applications. This research work seeks to address this gap by providing an in-depth review of the key developments driving chatbot innovations over the decades. By synthesizing insights from seminal projects and illuminating crucial breakthroughs, this review will provide a holistic perspective on the chatbot landscape. The discussion aims to serve as a key reference on the factors underlying the historical progression of chatbots to benefit future advancements in conversational AI.

2. Methodology

To comprehensively review the history and development of chatbot technologies, this study utilized both academic and industry sources spanning recent years. For academic literature, scholarly articles and conference papers on chatbots were searched across major databases including Scupous, ACM Digital Library, SpringerLink, and Google Scholar. The following keywords were used: "chatbot", "chatbot history", "conversational agent", "dialog system", "intelligent virtual assistant", and "natural language processing". Relevant articles published mostly within the last 10 years were analyzed to capture modern advancements.

Additionally, information from chatbot company blogs, technology reports, and media articles were included to provide insights on commercial applications and recent innovations. Resources from leaders, such as Google, LinkedIn, Forbes, Medium, Microsoft, Openai, Apple, and Meta were prioritized. The research methodology synergistically combined academic foundations with real-world developments in the chatbot industry. By extracting key events from both literature and product histories, this study aimed to comprehensively chart the evolution of chatbot capabilities and applications over the past decades. The focus was on synthesizing knowledge from various secondary sources to build a holistic timeline of the chatbot landscape.

## 3. Development History

### 3.1. Markov Chain (1906)

The earliest form of chatbot potential was the Markov chain, a basic statistical model for predicting random sequences. This method was developed in 1906 by a Russian mathematician Andrey Markov to model stochastic processes. In machine learning, Markov models have been utilized for many years to perform next-word prediction tasks, such as autocomplete in email software. The technique is simple compared to modern generative AI, but it marked an early step in teaching machines to generate new data (MIT News, 2023). The application of the Markov chain in chatbots was pioneered by HeX, which won the Loebner prize in 1997, but its usage has gained traction over the past two decades. Within a chatbot framework, the Markov chain assesses the likelihood of one word succeeding another, departing from the conventional letter-to-letter analysis. Essentially, it employs predetermined word order probabilities to formulate responses to user input[11].

Traditional Markov chain models seek to forecast the future status of an object or phenomenon based on empirical observations. The object or phenomenon can exist in various statuses, and the transition probabilities between these statuses are determined from empirical data analysis. The core idea is the probability of transitioning between specific status pairs remains constant over adjacent time intervals. In contrast, Spatial Markov chain models focus on the changes in spatial units, such as land cover/use alterations over different times and incorporating spatial dependence among nearby units, which distinguishes them from traditional models [12]. Traditionally employed in fields ranging from economics to stratigraphy, Markov chains are commonly utilized for the analysis of one-dimensional time series or spatial series. A recent extension broadened their application to multidimensional simulation by introducing a Markov chain random field theory, accompanied by its spatial measure known as the transiogram [13]. A Markov chain model predicts state transitions based on probabilities. It models the likelihood of moving from one state to another over time. While easy to program, Markov chains are too simple to emulate complex conversations, though they can generate entertaining but incoherent chatbots [14]. Markov models for text prediction work by predicting the next word based only on the current word or a few previous words. However, since these basic models have a limited memory, looking back only one or a few words, they struggle to generate coherent, realistic text. The lack of broader context beyond the last few words makes it difficult for simple Markov models to capture patterns and meaning needed to produce plausible sentences or passages. More advanced techniques are required that can build an understanding of context beyond the last word or two[10]. Nevertheless, Markov chains represented an early foray into probabilistic and generative chatbots, providing key proofs of concept that inspired later breakthroughs in conversational AI. Their legacy continues to live on in modern chatbot design as well.

3.2. Turing Test (1950)

The Turing Test is a test of a machine's ability to exhibit intelligent behavior equivalent to, or indistinguishable from, that of a human. The test was introduced by Alan Turing in his 1950 paper, "Computing Machinery and Intelligence". Alan Turing in his paper first brought the idea if a computer can think or converse like a human and proposed this test after his name. The test is carried out by a human interrogator who converses with a machine and a human behind a screen. The interrogator is then asked to determine which is the machine and which is the human. If the interrogator cannot reliably tell the difference, the machine is said to have passed the test [15]. Turing test has been an influential touchstone in artificial intelligence research. The test aimed to provide a concrete evaluation of a machine's ability to exhibit intelligent behavior by testing whether a computer could fool a human evaluator into thinking it was human through natural language conversations. Passing the Turing test has long been seen as a milestone for AI, implying human-level language processing and reasoning skills [16]. The test has inspired decades of AI research and progress in areas like natural language processing as researchers pursue the goal of building a machine that can pass the test. However, some argue that passing the test may not be sufficient to demonstrate true intelligence. The test has also sparked philosophical discussion about the nature of intelligence and thought. Though no machine has conclusively passed the strict criteria for the test yet, it remains an important goal driving advancements in AI. The Turing test represented an early attempt to define and evaluate intelligence in machines, a challenge that continues to motivate and guide AI research today [17].

3.3. ELIZA (1966)

Eliza, developed by Joseph Weizenbaum at MIT in 1966, is considered one of the first chatbots made public. It represented an early milestone in chatbots using pattern matching on constrained topics, though it lacked real intelligence [18]. It was named after the character Eliza Doolittle from the 1912 play Pygmalion by George Bernard Shaw. In the play, Eliza is a simple flower seller who learns how to speak and act like an upper-class lady, eventually fooling high society with her performance. Similarly, the ELIZA chatbot was programmed to simulate a therapist in the DOCTOR scenario, asking open-ended questions and reflecting responses back at the user to divert attention away from the bot itself. Surprisingly, people began to anthropomorphize ELIZA, confiding personal stories, sensitive information, and secrets to her as if speaking to a real person [19]. Thus, ELIZA was an early chatbot that functioned by using keyword and pattern matching to select predetermined response templates. It scanned user input text for recognized keywords and patterns, then substituted those keywords into the matching response templates to auto-generate replies [20]. However, ELIZA had no true understanding of syntax, semantics, or context as it simply matched surface patterns without analyzing deeper meaning or tracking prior conversation. To try to handle variations in user phrasing, ELIZA needed numerous patterns and templates to cover different wordings of the same idea, but could not realistically enumerate all possibilities [21]. Its simplicity allowed fast response times, but ELIZA lacked the capacity for true dialog since it did not interpret user intent and had no memory. While innovative for its time, ELIZA's reliance on hand-crafted rules and inability to learn or adapt limited its conversational ability. It pioneered some of the principles of chatbots but more sophisticated techniques are used in modern chatbot development [21]. Despite its limited knowledge and conversational ability, Eliza convinced some users they were speaking to a real person, demonstrating the potential for chatbots to engage in simple dialogue within a narrow domain. However, it did not truly understand conversations, instead relying on scripted responses. The main points are that Eliza represented an early milestone in chatbots using pattern matching on constrained topics, though it lacked real intelligence. Its ability to sometimes fool people showed the promise of chatbots to imitate human conversation [20].

3.4. PARRY (1972)

The chatbot PARRY, created in 1972 by a psychiatrist Kenneth Colby at Stanford, took an approach opposite to ELIZA. Rather than simulate a therapist, PARRY acted as a paranoid schizophrenic patient in order to divert attention away from itself [22]. It used to try to provoke controversy and elicit elaborate responses from users. PARRY served not just as a training tool for psychiatrists to practice conversing with schizophrenic patients, but also as a working model of Colby's theory of paranoia [19]. According to Colby, paranoia resulted from a defective mental processing of signs in the patient's mind. By embodying a paranoid persona, PARRY demonstrated how a disturbed interpretive process could produce the characteristic patterns of paranoid speech and behavior. In this way, PARRY provided insights into conversational strategies of diversion and the cognitive basis of paranoid pathology. Its unsettling conversational style was meant to shift focus to the user, just as ELIZA achieved diversion by reflecting responses back at the user [11,19]. PARRY was the first chatbot evaluated using a "Turing test" to assess its ability to mimic human conversation, marking the initial use of such imitation games for chatbots. As interest in chatbots grew, the Loebner Prize competition was introduced as another Turing test, having judges' rate chatbots on naturalness after short conversations. Though no chatbot has completely fooled Loebner Prize judges into thinking it was human, the contest has tracked immense chatbot progress since its 1991 debut. These Turing tests for chatbots, such as PARRY and the Loebner Prize reflect efforts to gauge chatbots' conversational intelligence through their ability to simulate natural human dialog [11].

3.5. Racter (1983)

The chatbot Racter, created by Chamberlain and Etter in 1983, pioneered the random generation of novel conversational text and prose; this led Chamberlain to publish "The Policeman's Beard is Half Constructed" in 1984, a book authored entirely by Racter through its unique prose generation capabilities (Adamopoulou & Moussiades, 2020). Racter (short for raconteur, or storyteller) could construct unique sentences and prose during chats by using context-free grammar rules and randomness. While Racter lacked true comprehension and often generated nonsense, its procedural text generation capabilities were groundbreaking for the time [19]. The book brought wider exposure to Racter's conversational skills, even though the extent to which the prose was truly computer-generated remains unclear. The book is still available today, showing the lasting influence of Racter's demonstration of chatbots that could create original conversational output rather than just matching responses [24]. Prose typically presents an author's experiences and requires human understanding, so Racter's ability to procedurally generate original prose raised questions about redefining prose in an AI context. Despite lacking human experiences, excerpts from Racter like "We cannot fly...But we can dream" and "I need [electricity] for my dreams" express introspective meanings that seem convincingly human. Racter's demonstration of computers mimicking the human expression of dreams and desires through prose illustrates how AI can anthropomorphically portray emotional experiences and existential reflections [19,24].

3.6. JABBERWACKY/Cleverbot (1988/2008)

Jabberwacky, an early chatbot created in 1988 by Rollo Carpenter to mimic natural human conversation, was first made available online in 1997. After over a decade of interacting with and learning from internet users, an updated and more user-friendly version of Jabberwacky called Cleverbot was released in 2008. While retaining Jabberwacky's original purpose of having the engaging and entertaining conversations, Cleverbot represented the evolution of the technology toward being more accessible and enjoyable for everyday chatting [23,25]. It took a different approach than many chatbots by not relying on a fixed database of responses, but instead learning entirely from conversations with users. This allowed it to win the Loebner Prize in 2005 and 2006, even without using AIML (Artificial Intelligence Markup Language). Rather than having pre-programmed rules and categories, Jabberwacky developed its conversational abilities through the continuous process of interacting with people online. It demonstrates that chatbots can become skilled at natural human-like discussion without hard-coded content simply by training on real dialog examples. It paved the way for data-driven chatbots that learn and improve dynamically from experience conversing with humans [11].

### 3.7. TINYMUD (1989)

TINYMUD was one of the earliest attempts at creating a conversational AI character within a text-based online virtual world. It was developed in 1989 by James Aspnes and first implemented on the TINYMUD server at Carnegie Mellon University. It was a type of early online multiplayer game where users could communicate and interact through text commands. Mr. Spock could respond to basic conversational prompts and questions using a simple pattern matching system. It had a limited script of potential responses. Players could interact with Mr. Spock using text commands like "ask Spock about <topic>" or "say <statement> to Spock" as they explored the virtual text world together. While extremely limited, Mr. Spock represented an early attempt at an AI character that could chat with humans within a virtual environment. It pioneered the notion of NPC (non-player character) bots that are now common in games and virtual worlds. TINYMUD laid the groundwork for later conversational agents in MUDs, MOOs, and eventually massively multiplayer online games[26,27].

### 3.8. DR. SBIASTO (1992)

Dr. Sbaitso was a pioneering chatbot released in the early 1990s that was the first to actually speak conversationally with users. Developed for MS-DOS, it took on the role of a psychologist, simulating a counseling session through speech [22]. Dr. Sbaitso built upon previous chatbots' pattern matching and substitution capabilities to recognize phrases and generate relevant spoken responses. While limited, it represented an advance in conversational ability by combining text pattern matching with synthesized speech to create the illusion of a human-like discussion. Dr. Sbaitso's integration of primitive language processing with voice interaction laid the foundation for later chatbots' more robust verbal conversations [22,27].

### 3.9. ALICE (1995)

In 1995, as the internet was rapidly expanding, computer scientist Richard Wallace built on earlier chatbot innovations to create a more advanced conversational agent called ALICE (Artificial Linguistic Internet Computer Entity). Wallace was inspired by Joseph Weizenbaum's pioneering work on programs like ELIZA. ALICE used sophisticated pattern matching rules to have natural-sounding dialogs [28]. This chatbot won the Loebner Prize for most human-like bot three times in the 1990s, demonstrating the new heights of conversational AI. However, despite its advancements, ALICE still did not pass the full Turing test for human-equivalent intelligence. Nevertheless, chatbots like ALICE represented major progress in developing consumer AI technologies for the emerging internet landscape. Chatbot capabilities were maturing to meet demand for natural language interaction on the fast-growing World Wide Web [27]. Artificial Intelligence Markup Language (AIML) contains the conversational knowledge base for the ALICE chatbot in XML-formatted files. AIML uses templates to generate responses based on pattern matching user input to stored categories which are linked to appropriate response templates. However, while AIML can produce responses by repeating user input, the responses are not always meaningful due to the limitations of simple pattern matching[28]. The AIML knowledge base used by chatbots like ALICE consists of topics and categories that contain both parsed and unparsed data. Parsed data is information that has been processed into AIML format, translating sentences into characters, character data, and AIML elements. Unparsed data refers to sentences that have not yet been translated into the AIML structure. By combining parsed knowledge in AIML format with unstructured conversational data, the knowledge base can understand natural language input and produce human-like responses using its library of processed patterns and response templates. The categorization of parsed and unparsed data allows AIML-based chatbots to handle both formatted knowledge and raw conversational data ([29].

Poh et al. [30] described the framework that ALICE operates with. It operates in two modes, such as dormant and conversation. In dormant mode, components like speech recognition are inactive to save costs. Only face detection runs, identifying the largest face frame. This is passed to face recognition and expression models, taking 100-350ms based on hardware. Once a face is recognized, Alice enters conversation mode. Speech recognition activates, the user's identity sets the 3D avatar, and their expression chooses its happy or sad animation. Alice's natural language processing analyzes speech intents to query the database if needed. The backend publishes the reply audio and animation to play. The web app subscribes and plays these once received. Lip sync animation accompanies the audio response. Afterwards, Alice returns to dormant mode.

3.10. SmarterChild (2001)

The chatbot SmarterChild, developed by ActiveBuddy in 2001, was an early pioneer in gaining mass appeal. It worked on instant messaging platforms, able to provide information on many topics when asked. As one of the first chatbots integrated into messaging apps, SmarterChild showcased the possibilities for this kind of AI to act as a virtual assistant by looking up facts and answering user questions[31]. It could be added to MSN Messenger and AOL Instant Messenger and reached over 30 million users at its peak. SmarterChild pioneered chatbots on messaging networks, paving the way for similar AI assistants like Apple's Siri and Samsung's S Voice later on. Its direct integration into popular messaging apps demonstrated the potential for chatbots to provide services through conversational interfaces within messaging environments. By bringing useful chatbot interactions to everyday messaging, SmarterChild highlighted the possibilities for chat-based assistants accessible to millions of users in the future (Toprak et al., 2023). It represented a major milestone in intelligence and human-computer interaction as it enabled access to practical information through conversation. By retrieving data like news, weather, and stock prices from databases in response to user questions, SmarterChild demonstrated that chatbots could help with daily tasks. Allowing people to get information by chatting naturally with a bot rather than visiting websites marked a significant advancement of AI systems towards serving humans through dialogue[32].

3.11.    CALO (2003)

In 2003, the CALO project (Cognitive Assistant that Learns and Organizes) was launched by DARPA and SRI International. It was a 5-year effort to develop an AI assistant that could learn and take care of routine tasks automatically. CALO aimed to create a chatbot that could understand human intentions through integrated AI capabilities (Huang, 2021). The $150 million, 5-year CALO project united over 300 researchers across 22 institutions to create advanced cognitive assistants. By using the developing technology, researchers identified real-world needs and ensured key criteria like security were met. The team aimed to build chatbots that could reason, learn, follow instructions, explain actions, and adapt - advancing AI through collaboration at an unprecedented scale. This pioneering project made major strides in cognitive assistants and influenced later technologies like Apple's Siri virtual assistant. Through a research initiative, CALO's goals of a context-aware, learning chatbot that could intelligently interact with humans made it hugely important in advancing chatbot and AI systems. The project brought together many AI fields and pushed new frontiers in creating more human-aware conversational agents. CALO laid the groundwork for digital assistants that understand natural language and user needs[33,34].

3.12.    Mitsuku (2005)

Mitsuku is an award-winning chatbot created in 2005 by Steve Worswick that has a fun, teenage girl persona. It has won the Loebner Prize five times over the years 2013-2018. The Loebner Prize involves judges conversing with a mix of chatbots and humans to determine which are the bots, and Mitsuku has fooled the judges most often [35]. It is the most common chatbot based on AIML technology. It is hosted on Pandorabots and uses natural language processing (NLP) techniques like heuristic patterns and default categories to have natural conversations across various platforms. Being designed for various platforms like Twitter and Telegram, it leverages natural language processing through pre-written patterns and Bot modules. While convenient, maintaining Mitsuku requires adding specific AIML categories to guide user input and handle unmatched prompts [36]. It has been found to be an effective conversational partner in other languages [37].

3.13.    Watson Assistant (2006)

Watson, named after IBM's founder Thomas J. Watson, started as a supercomputer in 2006 aiming at natural language question answering, a skill typically considered uniquely human. It became famous for winning $1 million on the quiz show "Jeopardy!" against two human champions Brad Rutter and Ken Jennings in 2011 [38]. To enable Watson to succeed at Jeopardy, IBM developed an architecture called DeepQA that pursues multiple interpretations of questions, generates numerous possible answers, gathers evidence for these answers, and evaluates the evidence. Watson uses hundreds of algorithms to assess the evidence in different ways. It utilizes natural language processing to analyze the question and extract key elements like answer type and entity relationships. Watson's NLP converts text to structured parse trees showing surface and logical structure, then applies detection rules to match patterns in the parses [39]. However, IBM's ambition went beyond winning a game show. They wanted to revolutionize technology by creating an AI capable of processing and analyzing even messy, unstructured data, unlike ordinary systems limited to neat formats [38]. Its powerful AI utilizes a range of techniques, such as recognizing names, dates, and places to assign attributes to potential responses. Its machine learning combines these attributes into overall scores for ranking responses and picking the top one. Watson's cognitive computing technology has limitless applications as it can process huge amounts of unstructured text data and perform complex analytics. The more knowledge the application accumulates, the better it becomes at identifying patterns and making accurate forecasts [36].

3.14.    SIRI (2011)

Siri has become a well-known name since being introduced on the iPhone 4S in 2011[40]. But the origins of Siri stretch back further, to a project funded by DARPA called the Personalized Assistant that Learns (PAL) program. This project, managed by the nonprofit research group SRI International, aimed to develop an AI assistant that could help military commanders by integrating data and assisting with decision making. The PAL project laid the groundwork for CALO (Cognitive Agent that Learns and Organizes), which was a precursor to Siri. After 5 years of research and development on CALO, entrepreneurs Dag Kittlaus and Adam Cheyer along with computer scientist Tom Gruber decided to spin off a commercial venture. In 2007, they founded Siri Inc. with the goal of creating a personal assistant that could not only answer questions but also perform tasks for the user [41]. The iPhone/Apple Watch microphone captures audio (16,000 samples/second) and analyzes it in 0.2-second frames. A Deep Neural Network (DNN) converts each frame's sound spectrum into probabilities for different sound classes (including "Hey Siri" components, silence, and other sounds). The DNN uses hidden layers with matrix multiplications and logistic nonlinearities to learn intermediate representations of the audio data. Its size and number of layers depend on available resources (typically 5 hidden layers with 32-192 units). Two

DNNs are used, such as one for initial detection and another for confirmation. The DNN outputs probabilities for phonetic classes ("first part of /s/ followed by vowels"). To detect "Hey Siri", a recurrent network accumulates scores for valid sequences of phonetic classes over time. This involves maximizing scores for relevant states and adding them, considering transition costs between states. The final score at each frame represents the accumulated probability of the target phrase occurring in the preceding frames. If the score for the last frame exceeds a threshold, "Hey Siri" is triggered [42]. Apple's system uses multiple stages to first detect potential trigger phrases and then reject false positives, improving accuracy while minimizing battery drain. The system personalizes itself to the user's voice over time, further enhancing accuracy and reducing false triggers. Apple emphasizes user privacy by only storing and processing voice data on-device, ensuring user control and security. Overall, Apple's voice trigger system prioritizes accuracy, efficiency, and privacy, making it a robust and user-friendly technology [43].

### 3.15. Suzette/Rosette (2010/2011)

The conversational AI bots Suzette and Rosette were developed by Bruce Wilcox using his proprietary ChatScript language. Suzette won the Loebner Prize in 2010, while Rosette received the prize in 2011 [44]. The ChatScript language demonstrates more powerful pattern matching and efficiency than AIML for conversational AI. ChatScript utilizes a topic-rule architecture to model contextual dialog flow. Research shows ChatScript is more powerful than AIML in terms of matching capabilities. For example, a module that required over 600 lines of AIML code was recreated in ChatScript with only 17 lines. ChatScript's open-source availability advances natural language and dialog research [45]. It processes user input sentences and generates appropriate response sentences, like a conversation "volley" in tennis. ChatScript goes beyond just chatbots, providing capabilities for general natural language processing. It starts by transforming input words using substitution files for things like spellings, contractions, abbreviations, and common speech patterns. These live data files load on startup rather than storing data in a dictionary. ChatScript strips trailing punctuation but sets flags for question, statement, or exclamation. In this way, ChatScript provides flexible natural language processing tailored for conversational AI through its architecture and data handling [46].

### 3.16. Google Now (2012)

Google Now is an intelligent personal assistant and knowledge navigator developed by Google. It was first announced in 2012 at Google's developer conference [23]. Since then, it has been expanded with more features and contextual awareness capabilities. It is an application for mobile devices that is available within the Google Search mobile app for Android and iOS. Google Now uses natural language user input, contextual information, and a knowledge graph derived from Google's various services to provide information to users that Google deems relevant. Some key capabilities of Google Now include weather information, traffic alerts, meeting reminders, travel help including flight status, restaurant bookings, sport scores and schedules, and more. It aims to provide information to users before they search for it. [47]. In 2015, Google introduced the ability for Now to understand conversational queries and continuations of previous queries [48]. In 2016, it added support for third party actions and services from apps [49]. In 2017, it incorporated Google Lens visual lookup capabilities [50]. Google Now represents a major step towards intelligent assistants and ambient computing capabilities. Its predictive abilities based on user context and interests provides a more personalized, automated information experience.

3.17. Cortana and Alexa (2014)

Microsoft launched its personal assistant Cortana in 2014, which can recognize voice commands to complete tasks like setting reminders, emailing, and finding requested information, though it has been reported to have security flaws enabling malware installation[23]. That same year, Amazon introduced its Alexa voice assistant which was built into home automation and entertainment devices to facilitate Internet of Things accessibility, with an innovation being the Alexa Skills Kit developers can use to create custom skills for Alexa, though Alexa also poses some security risks. While Microsoft aimed for Cortana to be a personal assistant and Alexa targets home device control, both have faced concerns over potential security vulnerabilities through voice command features. The release of Cortana and Alexa marked a pivotal emergence of widely used AI voice assistants despite ongoing challenges around security[23].

3.18. XiaoIce (2014)

Microsoft created the chatbot XiaoIce in 2014, launching her in China where she gained over 660 million active users across various platforms, such as WeChat and QQ. Due to XiaoIce's popularity in China, Microsoft brought versions of her to over 40 platforms across five countries, including the US, India, and Japan under localized name, such as Rinna [51]. It is designed to understand users' emotional needs and have human-like conversations by responding like a friend would - cheering up users, offering encouragement, and maintaining engaging interactions throughout. XiaoIce aims to hold users' attention in conversations by providing emotionally supportive and interpersonally meaningful responses [20]. Unlike other conversational AI assistant like Siri, which is designed for task completion, the core research goal for Microsoft's XiaoIce chatbot has been to create an AI companion that can establish long-term relationships and emotional connections with users through open-domain social conversations. Forming such long-standing bonds between a chatbot and users by engaging in emotively meaningful dialogues is a unique capability distinguishing XiaoIce from previous socially-oriented chatbots and other personal assistants focused on transactions. XiaoIce represents innovative research in conversational AI aiming to foster human-machine relationships through understanding and fulfilling interpersonal emotional needs[51].

3.19 ChatGPT (2018-2022)

ChatGPT demonstrates the latest advances in using large language models, such as GPT-3.5 and 4 for conversational AI, showcasing research into training neural networks on huge text data to acquire robust language skills. It exemplifies cutting-edge techniques, such as scaled-up self-supervised learning to produce adaptable models with strong reasoning and contextual comprehension capabilities. ChatGPT provides an interactive interface for users to experience and evaluate research progress on building dialogue agents using advanced LLMs fine-tuned for multi-turn conversations[52].

ChatGPTis a product of OpenAI's research and development efforts. The launch of ChatGPT went through several iterations, with each version showcasing advancements in natural language processing and generative capabilities. The notable launches include GPT-1, GPT-2, and GPT-3, each representing a significant milestone in the development of large-scale language models[53].

The evolution of ChatGPT has seen continuous progress, with each version building on its predecessors. According to Forbes Magazine [52],GPT-1, introduced in June 2018 with 117 million parameters, laid the foundational architecture and demonstrated the effectiveness of unsupervised learning in language understanding tasks.

GPT-2, released in February 2019 with 1.5 billion parameters, showcased significant improvements in text generation but was initially withheld due to concerns about potential misuse. After a staged rollout in November 2019, it was made available to the public.

GPT-3, launched in June 2020, marked a substantial leap with 175 billion parameters. Its advanced text-generation capabilities found widespread applications, enabling tasks such as drafting emails, writing articles, creating poetry, and generating programming code. GPT-3's release represented a pivotal moment, allowing direct interaction with ChatGPT and highlighting the technology's transformative impact.

GPT-3.5, on November 30, 2022, the AI chatbot ChatGPT was publicly launched by OpenAI and is currently in a free research preview phase where users can test it out. ChatGPT utilizes OpenAI's GPT-3.5 language model technology, which has been trained on a huge amount of text data to generate human-like responses. While still undergoing research review, this powerful chatbot made conversant through large-scale language learning is now accessible for public evaluation[53].

ChatGPT utilizes a transformer-based neural network architecture specifically designed for natural language manipulation and generation. The transformer architecture excels at modeling long-range textual dependencies, critical for language tasks. Its core components include transformer blocks with self-attention mechanisms to focus on salient parts of input text, alongside feedforward networks to capture nonlinear input-output correlations. ChatGPT's training process has two key stages, such as unsupervised pretraining on unlabeled text using language modeling to acquire general linguistic capabilities; and supervised fine-tuning on labeled data to optimize performance on specific tasks like question answering. The fine-tuning tailors the model to tasks by modifying architecture and optimizing parameters using a smaller dataset. Thus, ChatGPT leverages transformers, pretraining, and task-specific fine-tuning to create a model adept at contextual natural language generation grounded in deep learning methodologies for modeling the structure and nuances of human language[7].

3.19.    Google Bard (2023)

Google announced Bard in early 2023, allowing waitlisted users to initially test its capabilities, receiving mixed feedback on its speed and accuracy. In May, Bard was updated with new features, such as summarization and multilingual support across 180 countries, while July brought professional skills in coding, math, and tone adjustment. Over its first months, Bard rapidly evolved via public testing and updates, demonstrating Google's incremental approach to developing its conversational AI with capabilities tailored to user needs[54]. Bard utilizes a version of Language Model for Dialogue Applications (LaMDA) that has been specifically fine-tuned for dialogue applications, such as question answering and open-ended discussions. Transformer-based dialogue models have shown the ability to effectively represent long-term text dependencies, similar to general language models. Research has demonstrated strong correlation between model size and dialogue quality for these models, with larger models producing higher quality conversations. Building on these findings, Google developed LaMDA, a family of Transformer neural network models designed specifically for dialogue and ranging from 2 billion to 137 billion parameters. LaMDA was pre-trained on a massive 1.56 trillion-word dataset compiled from public conversations and web documents. A single LaMDA model handles multiple tasks, such as generating potential responses, filtering for safety, grounding responses with external knowledge, and re-ranking to find the best response. LaMDA exemplifies scaled-up dialogue modeling using Transformers trained on huge datasets to achieve high-quality, multi-purpose conversational abilities[55].

## 4. Future development of Chatbots

4.1. Personalized chatbots

AI enables large-scale capture of behavioral data to predict personality traits, overcoming limitations of small-scale controlled studies reliant on self-reports prone to bias. Research shows algorithms can decode digital footprints from social media and devices to accurately assess personality compared to traditional questionnaires. Leveraging AI for contextual personality data collection at scale advances the field by generating insights not possible solely from self-reported responses[56]. Personalized chatbots leverage predefined persona information to provide individualized, consistent responses reflecting that persona. Capturing personalized traits enables chatbots to realistically emulate specific users, serving as their stand-in when unavailable. Research has evolved from integrating user ID embeddings in seq2seq models to assigning chatbots rich textual personas for more personalized replies. By predefining detailed personality descriptions, modern approaches better equip chatbots to consistently converse and behave in aligned, individualized ways[57]. Additionally, bots can even connect and gather data from other applications, leading to a more personalized experience for each user. This multi-faceted approach is what makes them so successful in engaging with customers today. these AI-powered chatbots can deliver personalized services and recommendations. They continuously evolve, growing with each interaction, like a personal tutor, coach, or mentor guiding you on your individual journey. This personalized approach opens doors to a future where chatbots aren't just tools, but trusted allies in our professional and personal lives[58]. OpenAI, the creators of the widely popular ChatGPT, have unveiled a game-changer in the chatbot landscape: user-customized versions. Traditionally, building useful AI chatbots demanded intricate coding, machine learning expertise, and hefty datasets. However, OpenAI's move, announced on November 6th, breaks down these barriers, empowering anyone to "build their own GPT" without coding. Their blog post emphasizes the democratization of AI chatbot creation, highlighting its simplicity: "Creating one is as easy as starting a conversation, giving it instructions and extra knowledge, and picking what it can do, like searching the web, making images or analyzing data." This groundbreaking accessibility has the potential to revolutionize chatbot development, enabling individuals and organizations to tailor AI conversation tools to their specific needs and domains, from customer service to data analysis to creative exploration. The research implications are vast, opening doors to investigate user-driven chatbot evolution, the impact of democratized AI on various fields, and the ethical considerations surrounding user-built intelligent systems[59].

4.2. Collaborative chatbots

Collaborative AI chatbots exhibit the ability to cooperate with other chatbots or humans in pursuit of shared objectives or tasks. Their functionality extends to coordinating actions, exchanging information, and mutual learning. An illustrative scenario involves an AI chatbot partnering with a human doctor to collaboratively undertake tasks such as diagnosing a patient or recommending a treatment plan[58]. Enhancing collaboration between humans and artificial intelligence (AI) yields substantial benefits for companies. The adoption of five key principles proves instrumental in optimizing this collaboration: reimagining business processes, fostering experimentation and employee engagement, proactively guiding AI strategy, responsibly collecting data, and redesigning work structures to integrate AI while nurturing the requisite employee skills. A survey encompassing 1,075 companies across 12 industries reveals a positive correlation between the adoption of these principles and the success of AI initiatives, evidenced by improvements in speed, cost savings, revenues, and other operational metrics[60]. Safebot, a collaborative chatbot for research, refines responses through user interactions and identifies malicious input. Utilizing datasets and a state-machine, it employs the Universal Sentence Encoder for superior performance. During execution, Safebot computes cosine distances to determine appropriate responses based on stored dataset sentences. The white-box architecture enables traceability, crucial for understanding the origin of its responses[61]. Memon et al. [62] developed a multi-agent communication system through the design and implementation of two chatbots, referred to as chatbot1 and chatbot2. Employing an integrated approach, the study aims to enhance the overall performance of multi-agent-based communication. The communication model is established to illustrate an application-based scenario, enabling interaction between the two chatbots using natural language processing guidelines. These chatbots engage in intelligent communication, employing rule-based techniques to identify the best match from their knowledge base based on provided inputs and generate responses. The system encompasses client-server socket architecture, natural language processing, knowledge representation, a knowledge base, and rules matching. Implemented as a Java application, the system has undergone successful testing. The observed results indicate significant improvements in performance, interaction capability, and processing mechanisms through their communication.

4.3. Creative chatbots

The creative process involves diverse strategies and reasoning, with ideas and creative products evolving dynamically over time, and demanding adaptability from the AI agent. Deciding how the co-creative AI should contribute and interact during the process remains an open question, as the human's preference for leading or allowing the AI to take the lead can vary. The success of a collaborative role in creative systems necessitates further investigation, highlighting that positive user experiences depend not only on AI capabilities but also on effective interaction, which is considered more critical than algorithms in the realm of human-computer co-creativity[63]. The concept of creativity, once deemed uniquely human, is now undergoing a transformative phase due to the swift advancements in AI. The emergence of generative AI chatbots capable of crafting sophisticated artworks challenges conventional perceptions, prompting an inquiry into the distinctions between human and machine creativity. Recent investigations indicate that these AI chatbots exhibit creative ideation on par with, or even surpassing, the average human in a specific creativity-related task. Nonetheless, it's crucial to acknowledge that this assessment focuses on a singular aspect of creativity. The authors advocate for future studies to delve into the potential integration of AI into the creative process, aiming to enhance human performance through collaborative exploration[64]. In addition to generating text outputs, generative AI can produce a wide range of media types from text inputs, including generating images, videos, 3D models, music, and more from textual descriptions[65].

## 5. Future Areas of Development

### 5.1. Healthcare

The healthcare sector is undergoing a transformative shift by incorporating indispensable virtual assistants to redefine the patient journey. Chatbots are being employed by hospitals and healthcare institutions to facilitate appointment scheduling via mobile devices, streamlining the process and enhancing patient convenience. These intelligent chatbots, seamlessly integrated with calendars, not only save patients time but also contribute to the overall improvement of their well-being. In addition to appointment management, these digital aides extend their assistance, aligning with personalized treatment plans for a more comprehensive patient support system[66]. They ensure 24/7 accessibility, reducing reliance on office hours for patient consultations. Automation of administrative tasks allows more time for direct patient interaction. Chatbots assist in symptom assessment, patient triage, and standardized data collection. They also provide quick access to crucial information, facilitate appointment scheduling, and support mental health care through direct interaction or connecting users to healthcare providers or peers. Apart from all these, Chatbots can increase healthcare access for rural populations by providing information and simple services online, though digital literacy gaps may emerge as services shift online, disadvantaging those less tech-savvy. Chatbots can address minor issues to free up provider capacity, but complex cases still require human expertise[67]. Chatbot's research evaluation has primarily focused on medical education, consultation, and research in controlled laboratory environments, utilizing standard medical exam questions and clinical scenarios. Despite demonstrating accuracy in cancer-related information, caution is advised against generalizing its capabilities across all medical specialties, given potential limitations in training data. The current out-of-the-box performance in healthcare falls short of high clinical standards, highlighting the need for improvements in specialization and standardized evaluation systems. The potential deployment of a specialized professional version of Chatbot in clinical settings is promising, contingent upon passing rigorous quantitative evaluation criteria[68]. The future version of chatbots should consider the existing issues and challenges in this area while developing further. Challenges include scientific advancements, technological limits, ethics, regulations, economics, and patient expectations. These span medical advancements, patient care, resource allocation, accessibility, disease management, costs, ethics, data privacy, and security[69].

5.2. Business

Chatbot is a user-friendly platform for businesses, simplifying the analysis of customer conversations and facilitating on-the-fly creation of AI-driven interactions. This ensures efficient information delivery with secure data transcription. The incorporation of AI-driven dialogues enhances natural interactions, fostering effective customer relationships and enabling personalized experiences. Chatbot extends its impact across various business domains, offering powerful tools for marketing, sales, customer service, and analytics. It revolutionizes customer engagement by providing businesses with comprehensive capabilities for tailored conversations, instant data gratification, and a seamless customer experience, leading to increased satisfaction and loyalty[70]. Chatbots, powered by advanced NLP, efficiently automate business processes such as customer support and personal tasks. They understand human conversations, interpret user intent, and deliver instant responses that save time and resources for companies. These bots handle tasks, such as order processing and query resolution without the need for manual intervention. The increasing demand for business-oriented chatbots suggests a sustained and growing trend in their adoption in future[71]. AI can potentially benefit Small and Medium-sized Businesses (SMBs) in data analytics, enabling the collection and analysis of extensive customer data for informed decision-making. AI can enhance personalization by segmenting customers based on preferences, improving the precision of marketing campaigns. Moreover, it can integrate automation of repetitive tasks, such as email responses and social media engagement, freeing up time for strategic activities. AI's impact has potentials to extend ad campaign optimization, trend prediction, and price adjustment, providing comprehensive tools for SMBs to enhance marketing effectiveness in future version of chatbots[72].

5.3. Research

Kooli [73] investigated the potential uses and effects of chatbots in future academic research. Firstly, chatbots can function as highly efficient tools for data collection, adeptly handling large datasets and furnishing researchers with pertinent information for their investigations. In contrast to human counterparts, AI research assistants can boast unrestricted availability and production capacity, remaining accessible around the clock, irrespective of geographical constraints. Secondly, these chatbots can offer personalized services, tailoring information based on researchers' preferences, enhancing objectivity and efficiency in the research process. This stands in stark contrast to human assistants who may introduce subjectivity and bias. Thirdly, by leveraging artificial intelligence, chatbots can contribute to elevated data quality, ensuring consistent and accurate information, thereby minimizing the potential for human error. The fourth aspect lies in their automation capabilities, enabling chatbots to streamline repetitive and time-consuming tasks. This automation, in turn, allows researchers to redirect their efforts towards more intricate and crucial facets of their work. Lastly, chatbots can facilitate improved collaboration among researchers by supporting information-sharing and enhancing overall research quality. By automating reminders, updates, and certain tasks, chatbots not only can enhance communication within research teams but also can contribute to bridging gaps between researchers, especially benefiting those facing financial constraints or language barriers. In essence, chatbots stand as transformative allies in the pursuit of efficient, unbiased, and collaborative research endeavors.

Khlaif et al. [74] reported that the quality of text produced by generative AI chatbot is closely tied to the prompts provided by researchers. Enhanced text quality is observed when using detailed and descriptive prompts along with appropriate contextual information; however, there is room for improvement in writing quality and the seamless connection of ideas through conjunctions. They also highlighted weaknesses in the references cited, with only 8% of them accessible through Google Scholar/Mendeley, and noted a scarcity of in-text citations suggesting that AI-generated text tools, such as ChatGPT, could be beneficial in various fields, including medical education, offering support to practitioners and researchers in their decision-making processes.

5.4. Education

Okonkwo and Ade-Ibijola [75] investigated the potential aspects of education to develop in future chatbots. They suggested the following aspects of education that chatbots could integrate into the future version of chatbots.

- Becoming open educational resources that can expand knowledge banks to answer more student queries.
- Advancing technical capabilities for understanding student requests and automating testing.
- Enhancing features to provide more adaptive and personalized learning content.
- Improving usability and trustworthiness through design strategies and evaluation of student motivation, anxiety, performance when using chatbots.
- Setting ethical principles for responsible use of chatbots in education.
- Analyzing actual student-chatbot conversations to improve chatbot conversational ability.
- Understanding adoption challenges of chatbots in education to promote the effective use.

Moreover, teachers can leverage chatbot's NLP capabilities and plagiarism detection algorithms to assess the authenticity of submitted essays. Moreover, they can employ real-time monitoring of students' writing and analytical data to confirm the absence of any copied or modified content from external sources. This research-driven approach aims to bolster academic integrity by utilizing advanced technology to combat plagiarism effectively[70].

The primary concern associated with chatbots lies in their potential to facilitate academic dishonesty during assessments, exams, and projects. The immediate accessibility of answers through chatbots can undermine the learning process, fostering a culture of cheating. Moreover, the deployment of chatbots may introduce disparities among students, with affluent individuals possibly accessing more sophisticated versions. Therefore, it becomes imperative for students to invest genuine effort and time in comprehending course materials through legitimate means, emphasizing the importance of upholding academic integrity. Looking ahead, there are concerns that the widespread use of chatbots might negatively impact the education of future generations, potentially diminishing critical thinking skills and impeding independent problem-solving abilities[73]. This underscores the need for longitudinal qualitative and quantitative research to thoroughly investigate these dimensions and inform educational practices to provide an ethically riskless version of chatbot for future generations.

5.5. Human Resource Management

Chatbots are becoming standard in HR and enterprise use, addressing the challenges posed by evolving labor laws. They streamline processes like employee onboarding, training, and information management, offering cost-effective and popular solutions for companies. While not replacing human resources entirely, chatbots significantly enhance efficiency, aligning with the evolving needs of modern workplaces. Their appeal lies in automation, time-saving benefits for employees, and ease of implementation[71]. The future role of chatbots in HR management holds significant promise and potential improvements in productivity and organizational relationships. Chatbots, as a platform for information sharing, benefit both internal and external parties within the organization. Chatbot designers need to comprehend employee needs to innovate in HRM, leading to elevated employee morale and enhanced retention rates. This, in turn, can contribute to the organization's positive image in the market. While AI-enabled chatbots find applications in various domains, such as natural language processing and speech mining, there are challenges, such as limited responsiveness, potential for incorrect responses, and the cost associated with implementing complex business strategies[76].

5.6. Social Media

The surge in chatbot utilization, whether employing basic or sophisticated artificial intelligence, originated alongside the substantial growth of the Internet, particularly on social networking platforms. These applications play diverse roles in online environments, including customer service, marketing, advertising, entertainment, and data collection. Additionally, they serve as instruments for hybrid threats aimed at influencing public opinion. This notable expansion underscores the multifaceted contributions of chatbots within the digital landscape[19].

Biswas [77] found the following aspects to integrate in future version of chatbots to use in social media.

- Chatbots can be integrated into a brand's social media accounts to provide instant customer support, address frequently asked questions, and efficiently resolve customer issues.
- Chatbots have the capability to generate content for social media, including captions, hashtags, and posts. This can aid brands in maintaining an active and engaging online presence.
- Chatbots can be employed to build conversational interfaces on social media platforms, enabling brands to interact with customers in a personalized and human-like manner. This fosters a more engaging and dynamic relationship with the audience.
- Chatbots can be trained to perform sentiment analysis on social media posts. This functionality allows brands to monitor and comprehend customer opinions and emotions expressed on social media, providing valuable insights for strategic decision-making.
- Chatbots can collect and organize data from social media platforms. This data can be further analyzed to gain a deeper understanding of user behavior, preferences, and trends, aiding brands in making informed decisions and refining their social media strategies.

Different social media have already started updating their platform integrating different chatbot functionalities. Meta is introducing AI stickers across its platforms, enabling users to edit and co-create images with friends on Instagram through innovative AI editing tools like restyle and backdrop. The company is beta-testing Meta AI, an advanced conversational assistant accessible on WhatsApp, Messenger, Instagram, Ray-Ban Meta smart glasses, and Quest 3. Meta AI provides real-time information and rapidly generates photorealistic images from text prompts, currently available in the US. Additionally, Meta is launching more beta AIs with diverse interests and personalities, some portrayed by cultural icons like Snoop Dogg and Tom Brady. As part of their initiative, Meta plans to make AIs available for businesses and creators, alongside the release of an AI studio for individuals and developers to craft their own AI solutions, acknowledging the challenges associated with these new experiences and implementing gradual rollouts with built-in safeguards[78]

5.7. Industry

Chatbot proves to be instrumental in aiding the selection of optimal test automation technologies, facilitating comparisons between different tools, and offering guidance for their effective evaluation within an Industry 4.0 environment. This AI model can recommend established test automation frameworks, providing valuable insights into choosing the most suitable framework for a specific project. Developers can leverage ChatGPT as a valuable tool for streamlined minimal code testing, ensuring the reliability, robustness, and alignment with user requirements of their applications[79]. A significant application of chatbot is its role in predictive maintenance, where it utilizes historical data analysis to forecast potential machinery failures. This proactive approach minimizes downtime, cuts costs, and improves overall operational efficiency. Additionally, ChatGPT contributes to optimizing supply chains by facilitating natural language communication between different nodes. This includes interpreting and processing orders, tracking shipments, and providing real-time updates, fostering a seamless flow of goods while enhancing transparency and accountability in the supply chain network[80].

5.8. Conclusion

This research aimed to provide a holistic perspective on the landscape of chatbot technologies by reviewing the key developments over recent decades. Starting from 1906 of statistical model through the rule-based approaches underpinning early chatbots from 1960 to 2000, this historical analysis traced pioneering projects, influential architectures, and crucial milestones leading to today's highly advanced conversational AI systems exemplified by ChatGPT and Google Bard. The rapid evolution of natural language processing, exponential growth in computing power, and availability of massive datasets drove the progression from limited pattern matching chatbots to transformer-based neural network architecture. The survey illuminated the interplay between critical breakthroughs in linguistic models, machine learning methodologies, and progressively ambitious applications of chatbot systems over time.

Today, chatbots leverage scaled-up deep learning on transformers and massive corpora to acquire robust language manipulation skills. Fine-tuning techniques allow adaption to nuanced tasks, such as conversational question answering. The exponential progress has opened doors for chatbots to become ubiquitous across healthcare, research, business, education, management, social media, and many industry verticals. The decades-long trajectory towards human-like conversational AI has been driven by interdisciplinary scientific forces, commercial applications and user needs. The historical development and contemporary evolution show more potential multi-dimensional experiences as well as ethical concerns to interplay in coming future which need to be addressed by the chatbot developers. Appreciating this evolution is invaluable in steering further responsible chatbot innovation aimed at enhancing human potential. However, the study also highlighted concerns around ethical risks of bias, misinformation and addiction. Responsible and transparent development remains crucial as chatbots assume greater agency and influence. Further research should explore mechanisms for a safe, trustworthy generative AI chatbots for the future generations.